# FROM OHMIC TO BALLISTIC TRANSPORT IN ORIENTED GRAPHITE

N. García[1,2], P. Esquinazi[1,2,3], J. Barzola-Quiquia[1], B. Ming[4] and D. Spoddig[1]

1. Division of Superconductivity and Magnetism, Institut für Experimentelle Physik II, Universität Leipzig, Linnéstraße 5, D-04103 Leipzig, Germany
2. Laboratorio de Física de Sistemas Pequeños y Nanotecnología, Consejo Superior de Investigaciones Científicas, Serrano 144, E-28006 Madrid, Spain
3. CMAM, Universidad Autónoma de Madrid, Cantoblanco, E-28049 Madrid, Spain
4. Electromagnetic Engineering Lab., School of Electronic Information Engineering, Beijing University of Aeronautics and Astronautics, Beijing 100083, China

Abstract

In this work we show that the spreading ohmic resistance of a quasi-2D system of size $\Omega$, thickness $t$ with a constriction of size $W$ connecting two half-parts of resistivity $\rho$ goes as $(2\rho/\pi t)\ln(\Omega/W)$, diverging logarithmically with the size. Measurements in highly oriented pyrolytic graphite (HOPG) as well as numerical simulations confirm this relation. Furthermore, we present an experimental method that allows us to obtain the carriers' mean free path $\ell(T)$, the Fermi wavelength $\lambda(T)$ and the mobility $\mu(T)$ directly from experiments without adjustable parameters. Measuring the electrical resistance through microfabricated constrictions in HOPG and observing the transition from ohmic to ballistic regime we obtain that $0.2~\mu m \leq \ell \leq 10~\mu m$, $0.1~\mu m \leq \lambda \leq 2~\mu m$ and a mobility $5\times10^4~cm^2/Vs \leq \mu \leq 4\times10^7~cm^2/Vs$ when the temperature $T$ decreases from 270K to 3K. A comparison of these results with those from literature indicates that conventional, multiband Boltzmann-Drude approaches are inadequate for oriented graphite. The upper value obtained for the mobility is much larger than the mobility graphene samples of micrometer size can have.

## I. INTRODUCTION

If the mean free path $\ell$ is of the order or larger than the sample size, ballistic transport and not Ohm´s law rules the physical properties. Therefore, a requisite to understand the transport properties of micrometer size or smaller conducting samples is the knowledge of the mean free path. If this is of the order or larger than the sample size (or the macroscopic crystallite size within a polycrystalline sample) Boltzmann´s equations cannot be used to determine the electrical resistance. Nowadays graphite is a material that attracts the interest of the solid state community, partially because of its anomalous transport behaviour, as the ordinary magnetoresistance (OMR) [1], ferromagnetism [2] and because is simply composed by graphene layers, which perfection is thought to be much larger than in single graphene layers attached to dielectric substrates [3]. Additionally, measurements of the OMR in graphite samples as a function of temperature ($T$) and sample size [4] indicate that the mean free path should be of the order of microns or larger at lower $T$. Therefore, the determination of $\ell$ is crucial to understand the previous phenomena. Not only $\ell$ but also the Fermi wavelength $\lambda$ is of importance because it is directly related to the carriers density $n$. If $\lambda$ is of the order of the sample size we will have also electron diffraction effects. The carrier density in turn determines the screening length that indicates the metallic, semiconducting or insulating behaviour of the material.

In the proposed band structure for idealized translational invariant graphite [5] all the previous physical parameters are temperature dependent. For decades, the transport properties of graphite have been studied by measuring the resistance R, the OMR, the Hall effect, etc., and interpreted them using the so-called two band model within the usual Boltzmann-Drude approach [6]. This model consists in a four temperature ($T$) dependent free parameters, namely two mobilities and two carrier densities and the basic assumption of diffusive, ohmic behaviour of the conduction electrons. The determination of these parameters using this model implies fitting transport data with at least four free parameters. In clear contrast to this general practice we develop here an experimental method that allows obtaining these quantities directly from experiment without free parameters or arbitrary assumptions. The idea is to use constrictions built in a graphite sample.

The transport of carriers shifts from ohmic to ballistic when $\ell(T) > W$, the constriction width [7]. Measuring $R$ as a function of $W$ as well as of $T$ we can determine experimentally and in an unique way $\ell(T)$ observing the shift and change in slope of $R(T)$, from ohmic to a ballistic regime as $T$ reduces. To our knowledge this is a clear way to distinguish the ballistic from the ohmic resistance of the sample and to determine the Knudsen-Sharvin resistance [7,8]; this regime cross-over should be rather universal and be observed for all constrictions. The difference between the resistance of the samples with and without constriction gives us $\ell(T)$. We note that our experiment provides an average behaviour of the carriers. As our sample is HOPG we discuss the problem as a 2D system with a given number of layers. Our data demonstrate that the carriers in HOPG have very large $\ell(T)$ in agreement with OMR results but also a giant mobility, which rises to $4 \times 10^7$ cm$^2$/Vs at low temperatures.

The paper has five more sections. The next section describes shortly the existing theory for the resistance of constrictions in three dimensions (3D) following with the corresponding formulas for the 2D case, our original contribution, and the description of the experimental method we propose to obtain the mean free path directly from experiments without adjustable parameters. Section III provides a few experimental details. In section IV we show the main experimental results. The discussion and the comparison of the results with published data are given in section V. The main conclusions are shown in section VI.

## II. OHMIC AND BALLISTIC CONTRIBUTIONS TO THE RESISTANCE OF 3D AND 2D SYSTEMS WITH CONSTRICTIONS

In what follows we first review shortly the existing theory for 3D constrictions and then develop the behaviour for 2D since our experiments will be done in 2D constrictions. We used a free electron gas model for massive electrons that is consistent with a band structure for low carrier density and parabolic dispersion relation [5]. For higher densities the dispersion may become linear but we are interested in densities between $10^8$ and $10^{11}$ electron/cm$^2$, where the transport is ballistic with a large mobility and mean free path.

Consider the geometry depicted in Fig.1(a) for a constriction in a screen. Since Maxwell it is known that a constricting circular orifice of diameter $W$ in a thin, non-transparent screen of size L$_s$ produces a spreading resistance that is equal to the $\rho(T)/W$ if the ratio $\ell/W << 1$ ($\rho(T)$ is the resistivity of the material). However when this ratio increases there are two corrections to the Maxwell spreading resistance: i) the ohmic value is corrected by a factor of the order of unity as pointed by Wexler [9], ii) and more important, a dominant ballistic term appears when $\ell/W \gtrsim 1$. This was observed by Knudsen and Sharvin [7,9] and the value of $R$ reads [9,8]:

$$R_{3D} = 4\rho(T)\ \ell/3A + \gamma(\kappa)\rho(T)/W \quad , \tag{1}$$

where $A=\pi W^2/4$ is the area of the hole or constriction and $\gamma(\kappa)$ is a smooth varying geometrical factor that is unity when $\kappa=W/\ell$ tends to infinity, recovering the Maxwell resistance, and is 0.67 when $\kappa$ goes to zero [9]. In (1) the first and second terms of the rhs correspond with the Knudsen-Sharvin and ohmic resistances, respectively. The ohmic, spreading resistance in 3D can be estimated within a factor $2/\pi$ off from the exact Maxwell´s solution assuming a hemisphere in which the electric field $E(r) = J_{3D}(r)\ \rho(T)$. The radius $r$ is taken at the constriction middle point and $J_{3D}(r)$ is the current density equal to the total current $I$ divided by half of a sphere, i.e. $J_{3D}(r) = I/2\pi r^2$ assuming that due to symmetry the current is radial far away from the constriction. For a similar calculation in the 2D case we have that the spreading, ohmic resistance of a constriction separating two semi-infinite media, is given by:

$$R_{s,2D} = 2(\rho(T)/t)\frac{\int_{W/2}^{\Omega/2}(I/\pi r)dr}{I} = 2\rho(T)\ ln(\Omega/W)/\pi t \quad , \tag{2}$$

where $\Omega$ and $t$ are the width and thickness of the system, respectively. This result does not take into account size effects near the constriction where the assumed radial dependence for $J_{2D}(r)$ may have deviations.

Solving numerically the Maxwell equations for the 2D case we have calculated the current lines and potential drop in a sample with squared geometry $\Omega=L_s$ for a constriction of length $L=W=4$ µm, see Fig. 2(a), and for $W = 100$ µm $>> L = 1$ µm, see Fig. 2(b). It can be seen that for the first case and at a radius $r > 4$µm the current lines have a radial dependence outwards of the constriction. However, in the vicinity and within the constriction this is not true. In particular, in the constriction the lines are nearly parallel to it, i.e. the constriction behaves like a tube with resistance $\rho(T)L/tW$. In the example for a 100μm constriction we see that the potential drops mainly around the constriction, see Fig. 2(b). The potential drop as a function of the position in the sample at different constrictions widths $W$ can be seen in Fig. 2(c). The calculated resistances as a function of $W$ for $L=1$μm and 4μm are plotted in Fig. 3. The logarithmic dependence is given exactly by the 2D spreading resistance of Eq.(2). This is an interesting result, which indicates a logarithmic divergence with the system size $\Omega$ in contrast to the 3D-Maxwell result where this resistance is system-size independent when the system size is large, i.e. $\kappa \to \infty$. The deviation from the logarithm behaviour at small $W$ is due to the resistance $\rho(T)L/(Wt)$ of the tube itself. We have calculated and plotted this correction and the total ohmic resistance in Fig. 3 for $L=1$μm and 4μm. We obtain an excellent agreement between the numerical simulation data and the sum of the two resistances without any fitting parameters.

All above is valid for the diffusive case, i.e. when the ratio $\ell/W \ll 1$. When this is not the case the Knudsen-Sharvin ballistic contribution given by

$$R_{b,2D} = \frac{\pi \rho(T)\ell(T)}{4Wt} = \frac{h}{4e^2}\frac{1}{N_s}\frac{\lambda(T)}{2W} \; , \qquad (3)$$

has to be taken into account. The expression at the rhs of (3) is valid for graphite (factor 4 in the denominator instead of a factor of 2 for metals) because it has two pockets of carriers at the Fermi level; $N_s$ is the number of graphene layers of the sample. The constants $h$ and $e$ are the Planck constant and the electron charge. The total 2D resistance is:

$$R_{2D} = \frac{a\pi\rho(T)\ell(T)}{4Wt} + \frac{2a\rho(T)\ln(\Omega/W)}{\pi}\bigg|_{W\ll\Omega} + \frac{\rho(T)L}{Wt}, \qquad (4)$$

where we include a constant $a \sim 1/2$ that takes care of the influence of the sample shape and the topology of the electrodes position in the sample. It can be obtained from the measurement of the resistance as a function of $W$, as it is shown below, or through numerical simulations knowing the exact positions of the electrodes.

Equation (4) has two temperature dependent functions, namely $\ell(T)$ and $\rho(T)$. Since $\rho(T)$ is a non-simple function of $\ell(T)$, it is difficult to obtain from experiments without making assumptions and fittings that have always some arbitrariness. However (4) has a very interesting structure, note that all the terms are multiplied by $\rho(T)/t$. This quantity can be obtained from the measurements of the sample without constrictions. Then if we divide $R_{2D}$ by this quantity we obtain a normalized resistance:

$$R^*_{2D} = \frac{\ell(T)a\pi}{4W} + \frac{2a\ln(\Omega/W)}{\pi}\bigg|_{W\ll\Omega} + \frac{L}{W}. \qquad (5)$$

This is an illuminating result because it does not contains $\rho(T)$ any more and depends only on the function $\ell(T)$, which can be extracted directly from the experimentally determined $R^*_{2D}$ without free parameters. This 2D equation will be applied for constrictions done in a HOPG sample. It should be noted that one cannot obtain the carrier mass from transport problems because the relevant parameters are the mean free path and the Fermi wavelength; to have the mass is necessary to obtain the Fermi energy.

It is interesting to notice that Eq. (5) implies that $R^*_{2D}$ grows proportional to $\ell$. This is provocative and not only due to the 2D case, is it the same as in the Sharvin-Knudsen formula [7-9] for 3D. The apparent divergence is due to the fact that the resistance $R$ is divided by the resistivity that behaves as $1/\ell$. But this apparent "divergence" is in the beauty of the ballistic theory, because $\ell$ never tends to infinity, but as it is clear from our experiments, it saturates. The maximum possible size of $\ell$ is the size of the sample.

Note that we assumed specular reflection at the boundaries. This is correct if the Fermi wavelength is much larger than the atomic structure because then multiple scattering with the atomic roughness is completely negligible. The profile of the constriction does not show irregularities comparable in size to the Fermi wavelength. Note that in our case the Fermi wavelength is larger than ~ 100 nm, especially at low $T$ where the errors are small. For metals the Fermi wavelength is of the order of the atomic structure and the assumption of specular reflection is more doubtful (see Ref. [8] where this point is discussed). In what follows we describe the sample characteristics and the preparation of the constrictions following with the experimental results.

## III. EXPERIMENTAL DETAILS

A HOPG sample of ZYA grade (i.e., 0.4° rocking curve width) with lateral dimensions of 3 mm length and 1 mm width was glued with varnish on a glass substrate. By peeling with double side tape the thickness was reduced from 300 µm to ~ 33µm. The glass substrate was placed inside a multi-electrode chip. The electric contacts were done with Au-wires and conductive silver in the four-point van der Pauw arrangement, see Fig. 1(a), with $L_s/1.25 \approx \Omega \approx 0.6$ mm. The resistance measurements were performed with an AC Linear Research LR700 bridge with currents between 100nA and 10µA. The quantity $\rho(T)/t$ has been obtained using the expression $\rho(T)/t = \pi (R_1 + R_2) f / 2 \ln(2)$ where $R_{1,2}$ are the measured resistance interchanging the current and voltage electrodes and $f = 0.95$, using the measured resistance ratio [10].

Due to the different boundary conditions between the arrangement of the simulations (see Fig. 2) and the van der Pauw configuration used in the experiment we have a small correction to the first two terms of Eq. (5), which is included in the constant $a$ in Eq. (4). The constrictions were produced at the sample middle and between the electrodes using the sputter capabilities of a FEI NanoLab 200XT dual beam microscope (DBM). The Ga$^+$ ion beam of the DBM allowed a defined surface ablation. The ion current was selected upon the constriction width, using a maximum of 7 nA and a minimum of 50 pA with 30kV voltage and a beam overlap of 50%. During the cut the resistance was measured in situ continuously, allowing us to identify when the cut was through all the sample thickness as well as the current distribution through the sample thickness. To reduce Ga$^+$ damage and to obtain sharp edges the cuts were done in two steps. After the initial cut the ion current and the length of the cut were reduced. Three constrictions of length $L \approx 35$ µm, 3 µm and 2 µm and width $W = 35$ µm, 16 µm and 2 µm were produced. Figure 1(b) shows a DBM picture of the smallest constriction.

To characterize the size of the crystallites in the HOPG sample we used the electron backscattering diffraction (EBSD) option in the DBM. The measurement was performed with a commercially available device (Ametek-TSL). In this setup the sample under investigation was illuminated by the electron beam and the diffracted electrons were detected by a fluorescence screen and a digital camera. The TSL software was used to calculate the orientation of the HOPG surface as function of the electron beam position. Figure 1(c) shows an area of 200 x 110 µm$^2$ of the HOPG sample where the grain distribution is recognized by the in-plane orientation depicted by the (blue-green) colour distribution. Crystallites in the ~5…20 µm range of elongation are observed.

## IV. EXPERIMENTAL RESULTS

*(a)    Determination of the electron mean free path $\ell(T)$ and Fermi wavelength $\lambda(T)$*

The resistances as a function of temperature were measured from 270 K to ~2.5 K for the sample without and with three different constrictions. Figure 4 shows the experimental values for the 4 cases. These results show that the resistance $R$ increases reducing the constriction width $W$. At high enough $T > T_1$ so that $\ell(T_1) << W$ the results show that $R$ follows the logarithmic dependence on the constriction width $W$, see Eq.(2). Plotting $R$ vs. $W$ and using (4) with $L \cong W$, the experimentally determined ratio $\rho(250K)/t = 22.8$ mΩ, from the fit we obtain $a = 0.42$, see Fig. 3. Note that we plotted $R(T_1=250K)$ because at that temperature or above the required condition for $\ell(T_1)$ is satisfied. When we divide the curves obtained for the constrictions by the $\rho(T)/t$ curve we find a result that can be related to (5).

Notice also that the good agreement of the logarithmic dependence with experiment, see Fig. 3, indicates that the resistivity in the constriction has not been modified. This is not surprising because: (1) the used ion beam spot is around 20 nm and the smaller constriction is 2 microns, two orders of magnitude larger. Therefore the constriction boundaries practically do not contribute. (2) The grain boundaries depth or the thickness of the single crystalline regions are less than 0.1 µm thick (transmission electron microscope measurements done parallel to the graphene layers on similar HOPG samples) and therefore, in 30 microns thick sample one has at the constrictions and in average the same grain boundary density as in other parts of the sample. Since the constriction region is part of the bulk sample it is expected that for the smallest constriction size the resistivity should remain similar to that of the bulk sample.

The experimental values for the normalized resistances are presented in Fig. 5. We see that reducing $T$ there is a clear increase in the experimental $R^*_{2D}$, which is due to the cross-over between the ohmic, diffusive to a ballistic regime when the mean free path $\ell(T)$ grows, an expected result. From (5) we obtain $\ell(T)$ using the experimental $R^*_{2D}$. Using the data for the smallest constriction the obtained $\ell(T)$ is plotted in Fig. 6. It grows from a value of ~0.2 µm at T > 200 K (note that the error in $\ell$ at these temperatures is large) to 10 µm below 10K. This maximum value appears to be restricted by the size of the crystallites in our HOPG sample, see Fig.1(c), because the mean free path cannot be larger that the crystallites conforming the sample. We think that this is an important result and are not aware of any other method to obtain the mean free path of the carriers without using adjustable parameters and additional assumptions.

The same behaviour, i.e. the increase of $R^*$ lowering $T$ is observed for all values of $W$. The same sample with larger $W$ yields similar values of $\ell$ but, as expected, the relative increase in $R^*$ is much smaller for the other two constrictions because $W >> \ell$ and then in those constrictions the dynamic is mostly not purely ballistic. Therefore, the error in estimating $\ell$ is much larger for those large constrictions, as it is the case for the smallest constriction ($W = 2$ µm) at high temperatures $T > 200$ K (see Fig.6). Notice that the deduction of the ballistic term [7-9] is done for the case of a screen, i.e $L$ tending to zero.

With these data we can now obtain the Fermi wavelength $\lambda(T)$ using (3) without any adjustable parameter, see Fig. 6. The obtained values indicate that $\lambda(T)$ is a factor ~5 smaller than $\ell(T<10K)$ increasing to ~ 2 µm at low temperatures, see Fig. 6. These values are consistent with a Fermi wavelength determined basically by thermal electrons, i.e. if the Fermi energy $E_F \sim k_BT$ and $\lambda = h/(2m^*E_F)^{1/2}$ one obtains 0.1 µm $\leq \lambda \leq$ 1.3 µm for 300 K $\geq T \geq$ 2 K and assuming an effective mass $m^* = 0.005m$ ($m$ the free electron mass) as an intermediate value between the mass-less Dirac fermions and massive ones with $m^* \sim 0.02m$, well between the limiting values obtained from old and new Shubnikov-de Hass and de Hass-van Alphen oscillations measurements [6,11]. Due to the square root dependence, a factor 4 change in the effective mass will not change substantially the estimated Fermi wavelength. This result indicates that for an ideal graphite sample we would expect that the electronic density $n$ tends to zero for $T \to 0$, expected for an ideal semimetal. For thin samples, however, sample preparation introduces in general enough defects in the graphite layers that may increase substantially the electron density.

Taking into account possible multi-band contributions to the resistivity and the applicability of Eqs. (4) and (5) to graphite, we stress again that our experiment provides an average values of the parameters of the carriers. For example, the contributions of two bands can be splitted in the model using an effective resistivity as $1/\rho = \sigma_1 + \sigma_2$, where $\sigma_i$ is the conductivity of the carriers of the $i$-band. The measured $\rho$ is the effective one we use to divide the total resistance. The mean free path $\ell$ in Eq. (5) is also an effective one. In case there are large differences between the conductivities and mean free paths between the bands, then our method provides the values for the carriers with highest conductivity and largest mean free path.

*(b) Determination of the mobility µ(T)*

We can also calculate the carrier mobility without any adjustable parameter according to

$$\mu = (e/h)\, \lambda(T)\ \ell(T)\,. \tag{6}$$

The mobility $\mu$ increases from ~ $5\times10^4$ cm$^2$/Vs to ~ $4\times10^7$ cm$^2$/Vs from 270 K to below 10K, see Fig. 7. In the same figure we show the carrier density that decreases from ~$5\times10^{10}$ cm$^{-2}$ to ~$2\times10^8$ cm$^{-2}$.

**V. DISCUSSION AND IMPLICATION OF THE OBTAINED RESULTS**

*(a) Comparison with results obtained in HOPG using the two-three band model under the Boltzmann-Drude approach*

The physics underlying the results presented in this work is very different from that found in metals, because in metals the electronic densities as well as the Fermi wavelength are practically temperature independent. In comparison with metals, the mean free path in HOPG is very large. It is so large that for micrometer size samples one cannot use anymore Boltzmann equation to describe the transport properties because both are comparable. Recent experiments revealed that the OMR of small graphite samples and in HOPG samples with constrictions is largely reduced in comparison with larger samples [4]. This is an indication of the lack of validity of Boltzmann-Drude classical theory in graphite.

In general the magnetotransport properties of oriented graphite have been interpreted in terms of two (or three) band models under the basic assumption of ohmic behaviour using

Boltzmann-Drude's theory. It should be clear that if the mean free path is of the order of the grain size, the above mentioned assumptions loss their validity. Taking into account the obtained mean free path in this work as well as the fact that the magnetoresistance of oriented graphite is sample size dependent [4], we should have doubts about the interpretation of the transport properties based in those models. Moreover, the fact that macroscopic HOPG samples are non-homogeneous, composed by micrometer large regions (patches) with different electrical properties [12], very likely due to defects in the graphite lattice (i.e. a non-intrinsic origin), makes also doubtful the use of homogeneous multiband models to understand graphite transport properties. To provide an example of this failure, we compare below the mean free path, carrier density and mobility obtained in this work without any adjustable parameter with those obtained in Refs. [13,14] where a two (or three) band model was used to interpret the magnetoresponse of HOPG and obtain the above mentioned transport parameters and their temperature dependence.

In Fig. 6 we have plotted the obtained mean free path vs. temperature. The observed temperature dependence can be in general described assuming the independent contribution of different scattering mechanisms (Matthiessen's rule):

$$\ell(T) = (\ell_0^{-1} + (\ell_e(T))^{-1})^{-1}, \qquad (7)$$

where $\ell_0$ denotes the scattering of electrons with the boundaries of the crystals inside the sample and $\ell_e(T)$ is a $T$-dependent mean free path where different scattering mechanisms, i.e. electron-electron, electron-phonon, etc., may play a role and can be, in principle, written also as the sum of independent scattering rates. The obtained $T$-dependence of the mean free path can be well fitted up to ~200 K using (7) with $\ell_0 = 10$ µm and $\ell_e(T) = 5.63 \times 10^3 \ T^{-2}$ µm (with $T$ in K), see continuous line in Fig. 6. The $T^{-2}$ dependence of the mean free path suggests an electron-electron scattering mechanism. In general and in usual metals, due to the smallness of the prefactor $(k_BT/E_F)^2$ the electron-electron scattering is relatively small. However, in the case of graphite with a carrier density that is basically thermally activated (see Sec. IV) we have $k_BT \sim E_F$ and therefore this scattering mechanism should have a non negligible contribution because the prefactor is of the order of unity.

Electron-phonon scattering has been several times invoked in the past as the main one for electrons in graphite. In fact, this scattering mechanism has been thought to be the reason for the $T^1$ dependence obtained for the scattering rate in Ref.[13], for example. However, the assumptions used in Ref.[13] to obtain the scattering rate and its $T$-dependence seem to be inconsistent with our results. With the relaxation rate $\tau^{-1}(T) = 0.065 \ k_BT/\hbar$ obtained using a three-band model and Boltzmann-Drude theory [13] we might try to estimate the mean free path $\ell_e(T)$ and compare it with our results. If we assume a constant Fermi velocity, e.g. $v_F \sim 10^6$ m/s, the relaxation rate [13] can be translated as $\ell_e(T) = 1.18 \times 10^2 \ T^{-1}$ µm, which inside (7) gives the dashed line shown in Fig. 6. However, due to the large change in carrier density with $T$ and assuming a parabolic dispersion relation, the Fermi velocity may not be $T$-independent. In this case and if one estimates $v_F$ using the experimentally determined Fermi wavelength with a constant effective mass m*, then the $T$-dependence of $\ell_e(T)$, using the scattering rate obtained from fitting magnetotransport data with the multiband model [13], would be even weaker than the one shown by the dashed line in Fig. 6.

On the other hand and due to the huge anisotropy of graphite (resistance ratio between normal-to and within-layers $\gtrsim 10^5$ [1]) we expect a highly anisotropic Fermi surface, actually a 2D Fermi surface as several experimental facts suggest [3]. In this case the scattering rate due to electron-phonon interaction in graphite may follow a weaker dependence, i.e. $T^2$ or $T^3$, than the usual $T^5$ at $T < \Theta_D$ (Debye temperature), as is the case for bismuth [15,16] or antimony [17].

In Ref.[13] the carrier density in 3D was obtained applying a three-band model to the HOPG transport data. This density is depicted in 2D in Fig. 7 (close stars). The actual carrier density of HOPG is between 10 and 200 times smaller than the one obtained in [13]. The mobility of HOPG obtained in Ref [14] within a similar formalism is also shown in Fig. 7 (open stars) and shows similar differences with respect to that obtained without adjustable parameters.

*(b) Comparison with graphene*

Equation (6) for the mobility is highly relevant. It indicates that for a 1 μm sample size, as those used to study graphene for example, the mobility can never be larger than ~$5 \times 10^5$cm$^2$/Vs because $\lambda(T) < 0.1$ μm to have at least 100-1000 electrons in the sample (i.e., n ~ $10^{10}$... $10^{11}$ cm$^{-2}$) with a maximum mean free path of 1μm. Recent results on suspended graphene show that the best obtainable mobility is ~$2 \times 10^5$cm$^2$/Vs [18,19]. This value may be increased by a factor of two or three. However, the reached carrier densities are still far away from the Dirac point. To decrease the electronic density having large mobility as in the case of HOPG the suspended graphene samples should have ~100 times larger areas (with similar or smaller defect density), actually a complicated issue nowadays.

Within our experimental arrangement and sample thickness we cannot change systematically the carrier density applying an electric field. However, because in HOPG the carrier density is basically thermally activated and it changes several orders of magnitude we can correlate the obtained mobility with the carrier density and compare it with data obtained for suspended graphene. This is shown in Fig. 8 with our data (■) and the data from Refs. [18,19]. From this comparison it becomes clear that the mobility values for HOPG are larger at similar temperatures than for suspended graphene samples and that the graphene layers inside HOPG have a much smaller carrier density, i.e. reach much closer the Dirac point than single graphene, a fact that speaks out for the higher perfection of the graphene layers in HOPG.

*(c) How can graphite show such a low resistivity ?*

There is an additional interesting point concerning the behaviour of the resistivity of graphite. For the sample studied in this work, the resistance as a function of *T* reduces a factor of 3 between 100K and 2.5K; it has a clear cross-over at ~100K, remaining nearly constant above this *T*. This in principle maybe consider not so provocative. However, note that in the same temperature range $\lambda(T)$ increases a factor of ~5 and the carrier concentration $n(T)$, which is proportional to $\lambda(T)^{-2}$, decreases a factor of ~25. If we consider that $\rho(T) \propto n(T)^{-1}$, then we would expect that it should increase actually a factor of 25 below 100K, instead of the measured decrease of a factor 3. In other words, we have a resistivity that decreases effectively a factor of ~75 and rather sharply below 100K. Additionally, we should ask ourselves, how it is possible that a semimetal with a carrier density of the order of $10^8$ cm$^{-2}$ can have such a low resistivity, i.e. $\rho \leq 10$ μΩcm at low *T* ? This is a very peculiar behaviour and points to superconductivity, representing additional hints to the recently done magnetoresistive experiments performed in HOPG [20]. In this experiment hysteresis in the magnetoresistance has been observed that can be only explained by Josephson coupling between superconducting regions through normal region, as well as magnetoresistance oscillations suggesting Andreev´s reflections. In this sense when we apply a magnetic field of 0.2 T perpendicular to the graphene layers the resistance of the sample studied in this work increases almost one order of magnitude at low temperatures, showing the usual metal insulator transition of graphite, previously reported and assigned to a possible superconducting state [1]. It remains to be seen whether pure bismuth, which shows a very similar magnetic field induced metal-insulator transition and also has a very low resistivity [13], a similar phenomenon may play a role. We note that bismuth, as graphite, has a low density, low effective mass of carriers and huge values of the electron mean free path [15,16].

## VI. CONCLUSION

The main messages of our work are:
- We prove experimentally and theoretically that the spreading ohmic resistance of a quasi-two dimensional conducting material with a constriction diverges logarithmically with the system and constriction size. This result overwhelms the Maxwell old result for three dimensional constrictions generally used. It is of general importance due to the broad application of this kind of formula to calculate the resistance of metals and semiconductors with constrictions and the resistance of contacts between them.
- We present an experimental method and the theoretical basis to obtain the mean free path and two other important transport properties of the transport carriers without adjustable parameters. The method is based on the measurement of the electrical resistance as a function of temperature through constrictions and observing the transition from the usual ohmic to the ballistic regime. To our knowledge this method is a free-parameters experimental method that allows obtaining the above mentioned

- quantities without fittings and further arbitrary assumptions. As an example we apply this method to oriented graphite.
- We obtain that the mean free path and Fermi wave length reach micrometer length at low temperatures in graphite. The mean free path is so long that it is limited by the crystallite size inside the sample (~10 micrometers) and therefore ballistic electronics in graphite can be a reality. Because of the huge values of these two parameters, experiments testing the conduction-electron diffraction phenomena in graphite may be possible.
- From the original data obtained in graphite and a further comparison with data for graphene we conclude that the huge mobility we obtain in our macroscopic graphite sample cannot be obtained in graphene or graphite samples of micrometer size due to sample size limitations. In particular we obtain 10 µm mean free path and a mobility of $4 \times 10^7$ cm$^2$/Vs (one of the largest ever measured for two-dimensional systems) in the ballistic region of graphite when the electronic density goes to zero (~ $10^8$ electrons/cm$^2$) below 10K. This is possible to obtain only in nearly perfect samples and without impurities (< 1ppm). Our results obtained in graphite as well as the above mentioned intrinsic limitation have a direct influence in the nowadays research activities on this kind of material (graphene as well as graphite). These results open new perspectives for ballistic electronics based on graphite.
- Finally, the results presented here as well as those already discussed in Ref. [4] cast strong doubts about the validity of the multiband Boltzmann-Drudde approach to understand the magnetotransport properties of oriented graphite and it needs further consideration.

**Acknowledgements:** This work was done with the support of a HBFG-grant no. 036-371, the Spanish CACyT and Ministerio de Educación y Ciencia, the Deutsche Forschungsgesselschaft under DFG ES 86/16-1, the European Union project "Ferrocarbon" and the DAAD under D/07/13369 "Acciones Integradas Hispano-Alemanas".

**Figure captions**

Fig.1(a): Sketch of the sample geometry with the electrode positions at the edges. The length and width of the constriction are shown as $L$ and $W$. (b): Scanning electron microscope picture of the 2µm constriction done on the HOPG sample. (c): Electron backscattering image shows the in plane orientation of the crystallites of the HOPG sample in a scan size of 200 x 110 µm$^2$; the crystallites observed here are in the ~10...20 µm range of elongation.

Fig. 2(a): Two dimensional plot of the current density distribution through a constriction of 4µm width, 4µm length. The thickness of the sample is 100nm. The colour scale represents the current density and the stream lines indicate the current flow with the density proportional to the current density too. (b) Two dimensional plot of the electric potential distribution and the current flow through a constriction of 100µm width and 4µm length. The colour scale represents the electric potential drop from 1V to 0V and the stream lines indicate the current flow inserted with arrows which sizes are proportional to the local current density. These numerical simulation are done for the case of the resistivity of Cu at room temperature, $\rho = 1.66$ µΩ.cm and 100nm thickness. The result scales for any other material and thickness. (c) Numerical results for the voltage across the sample length ($L_s = \Omega = 1000$µm) for different constriction widths $W = 1, 4, 10, 20, 50, 100, 500$ and $1000$ µm starting at the black continuous curve, and constriction length $L = 1$µm.

Fig.3: Resistance versus constriction width $W$. Left y-axis: The symbols (+, ) correspond to the numerical values of the of a Cu plate of 100 nm thickness with 1mm length and width and for two constrictions with $L = 1$ µm (+) and 4 µm ( ). The continuous (red) and dash-dotted lines correspond to Eq. (4) for the two constriction lengths. The parameters used are $\rho/t$=0.16x10$^{-3}$ Ω, which corresponds to Cu, $a = 1$ for a linear electrode arrangement, mean free path $\ell \ll W$. Right y-axis: (∗) Experimental values of the measured resistance vs. constriction width at $T = 250$K for the HOPG sample. The continuous line follows Eq.(4) with $\rho/t$=22.8mΩ, $a = 0.42$ and $L = W$ (note that the fit takes care only of the points where this relation holds).

Fig. 4: Resistance as a function of temperature for the HOPG sample without (+) and with three constrictions of different widths $W$. The plotted resistance for the sample without constriction is obtained as the average value of $R_1$ and $R_2$, interchanging the current and voltage electrodes, following Ref. [10].

Fig. 5: Ratio between the resistance $R$ of the HOPG sample with constriction and the resistivity ρ divided by the thickness $t$. The three curves correspond to the measurements for the sample with the three different constriction width W.

Fig. 6: Mean free path $\ell$ (■) and the Fermi wavelength λ (o) in microns, obtained from Eqs. (5) and (3), respectively, using the curve shown in Fig. 5 for $W = L = 2$ µm, the constant $a = 0.42$ and the measured ratio $\Omega/W = 300$. The error bars are calculated assuming an error of 10% in the ratio $\Omega/W$. The continuous line is a fit to the data using Eq. (7) and a $T^{-2}$ dependence for the mean free path $\ell_e(T)$. The dashed line is calculated using (7) and the scattering rate obtained by fitting with a three band model the magnetotransport data of HOPG as described in Ref. [12] assuming a $T$-independent Fermi velocity (see text).

Fig. 7: Electron density $n$ (■, left y-axis) and mobility µ (o, right y-axis) vs. temperature for HOPG. The close stars are the values of $n$ from Ref. [12] and the open stars the mobility values from Ref. [13], both parameters obtained using multiband models and the Boltzmann-Drude approach.

Fig.8: Mobility vs. carrier density obtained for HOPG (our data, (■) ). Note that the carrier density in our case changes with temperature that is why two temperature regions are depicted in the figure as a guide. The shadowed region called "Ballistic" is obtained from the ballistic model prediction at 100K from Ref.[18]. The suspended graphene data are obtained from Ref.[18] (shadowed area, and dashed line) for temperatures between 20 K and 300 K and from Ref. [17] (vertical error bars) at 5K.

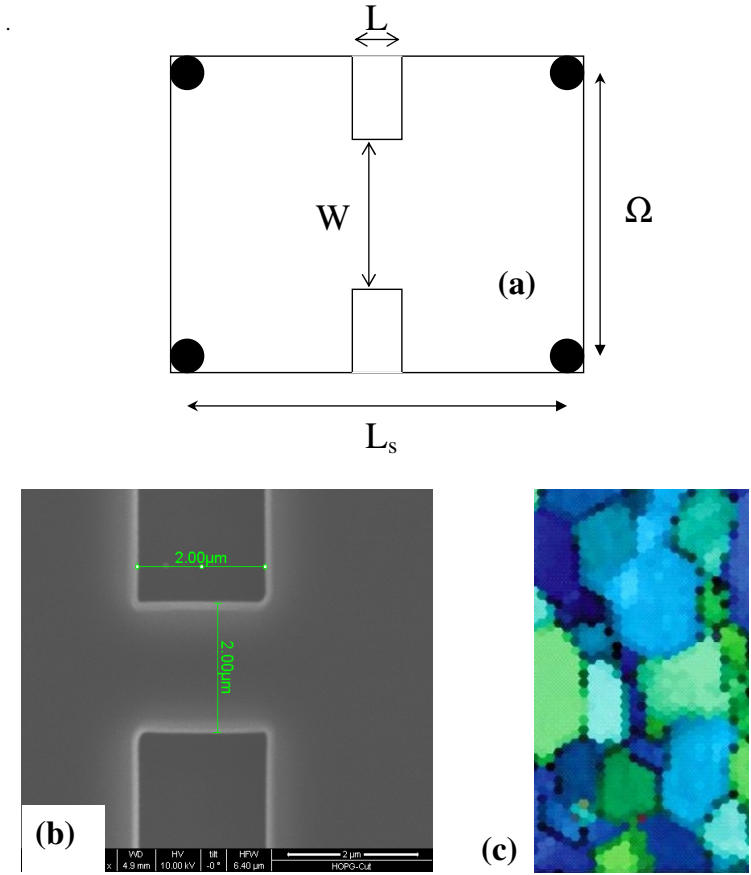

Figure 1

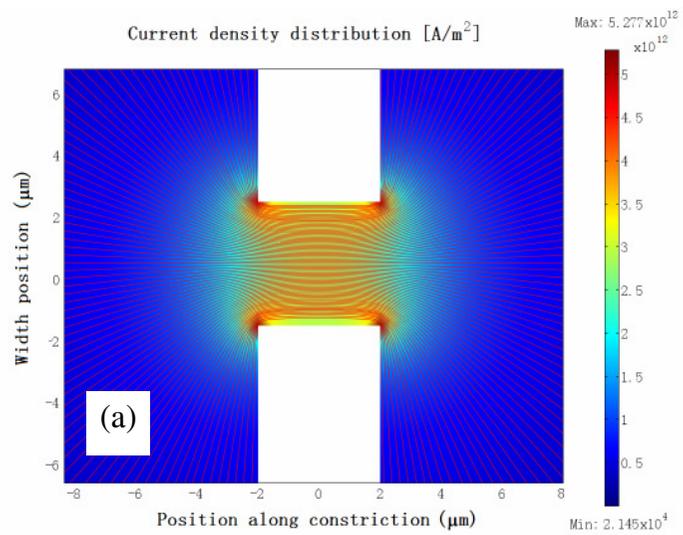

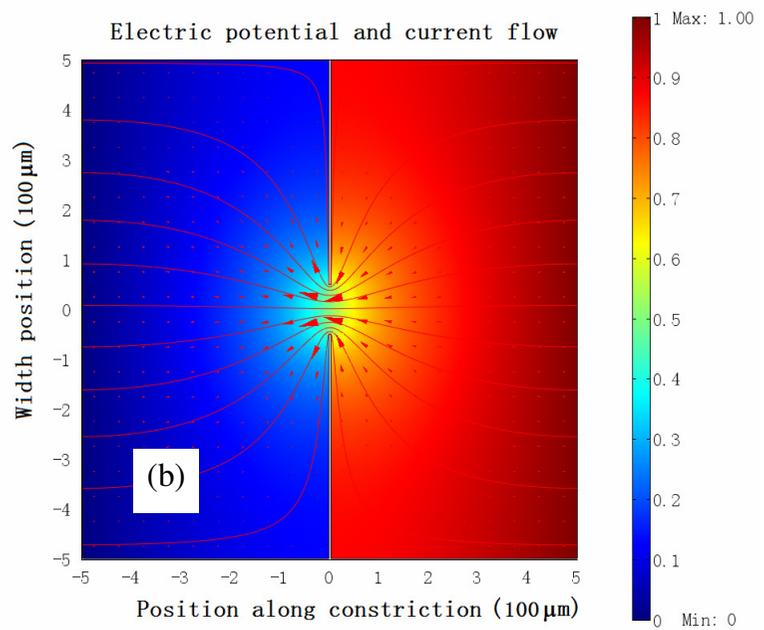

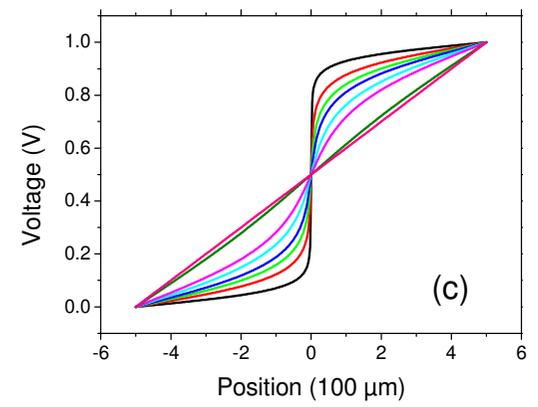

Figure 2

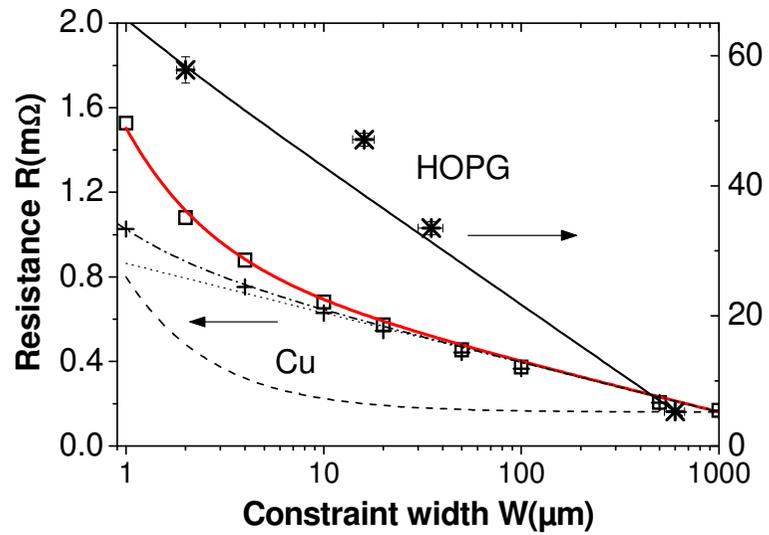

Figure 3

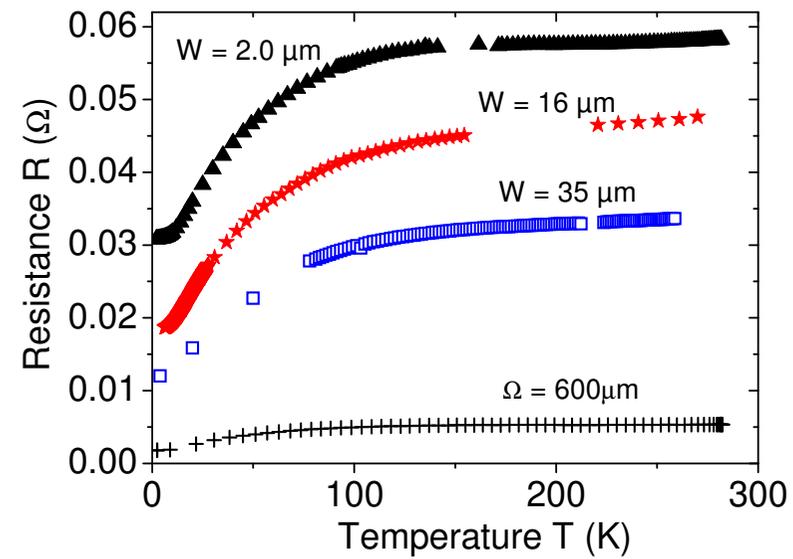

Figure 4

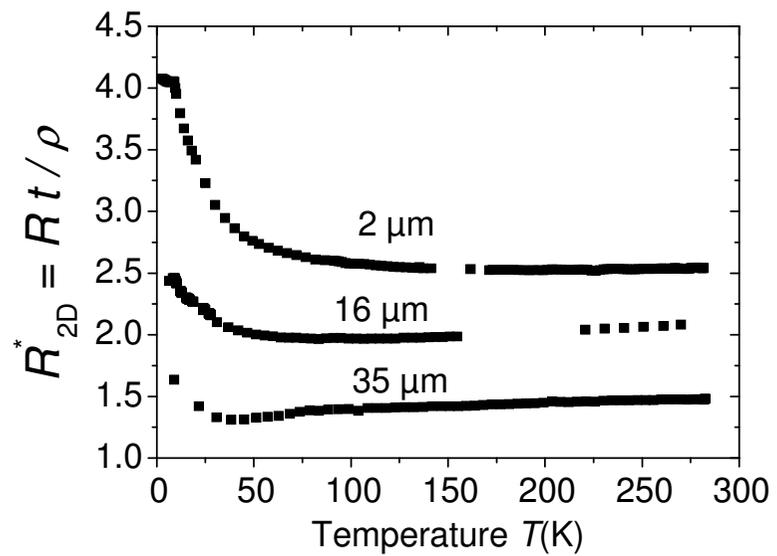

Figure 5

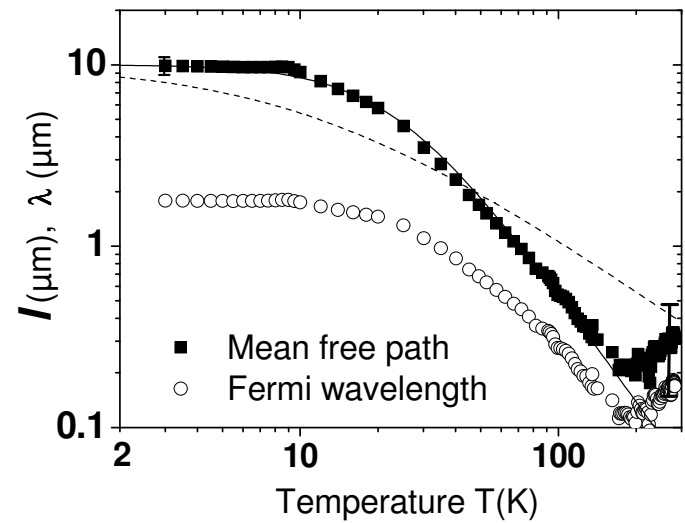

Figure 6

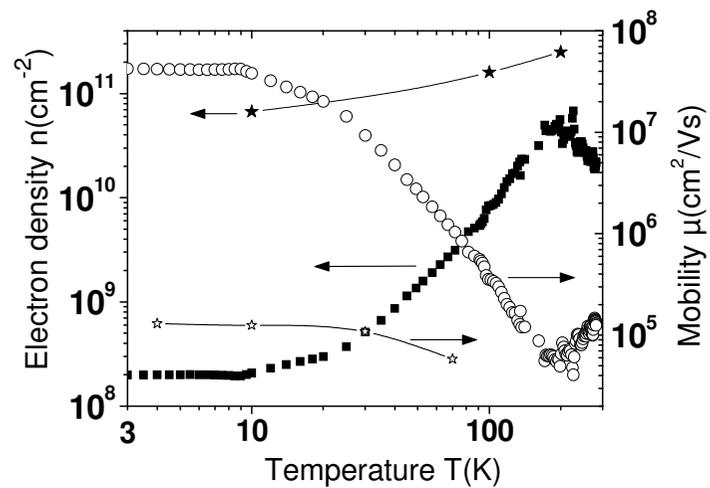

Figure 7

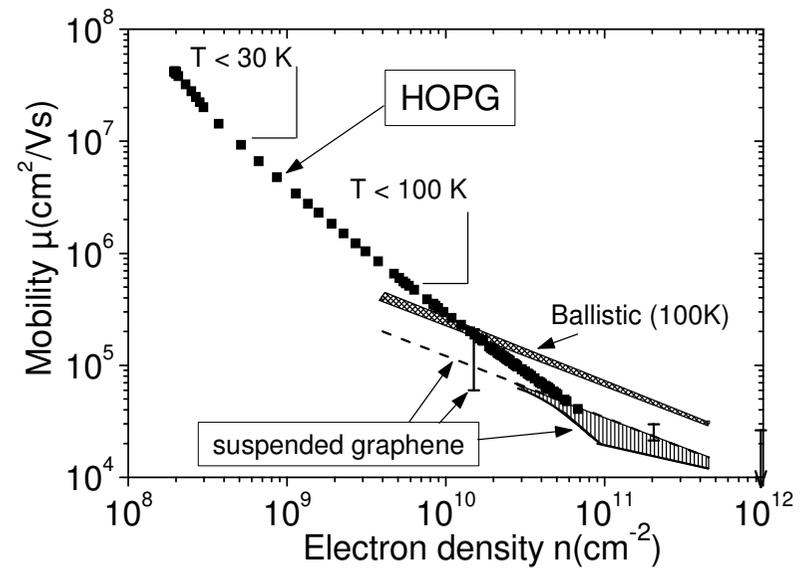

Figure 8